\documentclass[10pt,sigconf,letterpaper,review]{acmart}

\usepackage[framemethod=tikz]{mdframed}
\usepackage{soul}
\usepackage{enumitem}
\usepackage[hyphenbreaks]{breakurl}
\usepackage{makecell}

\AtBeginDocument{%
  \providecommand\BibTeX{{%
    \normalfont B\kern-0.5em{\scshape i\kern-0.25em b}\kern-0.8em\TeX}}}

\setcopyright{acmcopyright}
\copyrightyear{1999}
\acmYear{2021}
\acmDOI{10.1145/1122445.1122456}

\acmConference[Redacted Conference]{Redacted Conference 2063}{5 April 2063}{Earth, Alpha Quadrant}
\acmBooktitle{Proceedings of the Redacted Conference}
\acmPrice{15.00}
\acmISBN{978-1-4503-70899}

\usepackage{totcount}
\newtotcounter{finding}
\regtotcounter{finding}
\newenvironment{finding}[1][]{\refstepcounter{finding}\par
   \textbf{Finding~\thefinding:} \textit{#1.}\rmfamily}{}

   \renewcommand\footnotetextcopyrightpermission[1]{}
\begin{document}

\title{Unraveling Threat Intelligence Through the Lens of Malicious URL Campaigns}

\author{Mahathir Almashor}
\email{mahathir.almashor@data61.csiro.au}
\authornote{Commonwealth Science and Industrial Research Organisation (CSIRO) Data61}
\authornote{Cyber Security Cooperative Research Centre}
\affiliation{}

\author{Ejaz Ahmed}
\email{ejaz.ahmed@data61.csiro.au}
\authornotemark[1]
\affiliation{}

\author{Benjamin Pick}
\email{benjamin.pick@data61.csiro.au}
\authornotemark[1]\authornotemark[2]
\affiliation{}

\author{Jason Xue}
\email{jason.xue@data61.csiro.au}
\authornotemark[1]
\affiliation{}

\author{Sharif Abuadbba}
\email{sharif.abuadbba@data61.csiro.au}
\authornotemark[1]\authornotemark[2]
\affiliation{}

\author{Raj Gaire}
\email{raj.gaire@data61.csiro.au}
\authornotemark[1]\authornotemark[2]
\affiliation{}

\author{Shuo Wang}
\email{shuo.wang@data61.csiro.au}
\authornotemark[1]\authornotemark[2]
\affiliation{}

\author{Seyit Camtepe}
\email{seyit.camtepe@data61.csiro.au}
\authornotemark[1]\authornotemark[2]
\affiliation{}

\author{Surya Nepal}
\email{surya.nepal@data61.csiro.au}
\authornotemark[1]\authornotemark[2]
\affiliation{}

\begin{abstract}
The daily deluge of alerts is a sombre reality for Security Operations Centre (SOC) personnel worldwide. They are at the forefront of an organisation's cybersecurity infrastructure, and face the unenviable task of prioritising threats amongst a flood of abstruse alerts triggered by their Security Information and Event Management (SIEM) systems. URLs found within malicious communications form the bulk of such alerts, and pinpointing pertinent patterns within them allows teams to rapidly deescalate potential or extant threats. This need for vigilance has been traditionally filled with machine-learning based log analysis tools and anomaly detection concepts. To sidestep machine learning approaches, we instead propose to analyse suspicious URLs from SIEM alerts via the perspective of malicious URL \textit{campaigns}. By first grouping URLs within 311M records gathered from VirusTotal into 2.6M suspicious clusters, we thereafter discovered 77.8K malicious campaigns. Corroborating our suspicions, we found 9.9M unique URLs attributable to 18.3K multi-URL campaigns, and that  worryingly, only 2.97\% of campaigns were found by security vendors. We also confer insights on evasive tactics such as ever lengthier URLs and more diverse domain names, with selected case studies exposing other adversarial techniques. By characterising the concerted campaigns driving these URL alerts, we hope to inform SOC teams of current threat trends, and thus arm them with better threat intelligence.

\end{abstract}

\pagestyle{plain}



\maketitle

\section{Introduction}\label{sec:intro}

Timely and decisive threat intelligence is a vital imperative for cyber-security personnel defending their organisations. Within their respective Security Operations Centres (SOC)~\cite{kokulu_matched_2019, islam_multi-vocal_2019}, human teams are inundated with alerts for every potential incursion, and are often under great pressure to rapidly recognize and react to high priority threats~\cite{rosso_saibersoc_2020, akinrolabu_challenge_2018, sundaramurthy_tale_2014}.

Security Information and Event Management (SIEM) \cite{fireeye_what_2022, cofense_siem_2022} systems present in most SOCs collect and compile real-time events from an organisation's end-users, applications and devices \cite{najafi_malrank_2019}. These are then surfaced to SOC analysts to sift through and verify, with good threat intelligence playing an outsize role in identification and prioritisation efforts \cite{shu_threat_2018}. Machine learning based logs analysis \cite{ban_combat_2021, oprea2018made}, data aggregation and inference \cite{merah_ontology-based_2021}, and anomaly detection \cite{ahmed_survey_2016} all aim to lower the burden for human operators in such settings.

Aforementioned approaches can be improved since we cannot simply rely on a huge swath of Uniform Resource Locators (URLs) classified into benign or malicious by deep learning. For learning based detectors, threat intelligence such as this could help tease out features and connections amongst disparate URLs. To this end, we propose a salient alternative that focuses on bettering threat intelligence through the lens of malicious URL campaigns. SOC teams are inundated with suspicious URLs, be it from communication channels (e.g., within a phishing email), or from application and network events (e.g., a compromised device phoning home). Malicious actors exploit this weakness by incorporating URLs into multi-stage strategies, with a reliance on waves of diverse URLs to maximise reach and evasion. We witness the recent surge of attacks based on the COVID-19 pandemic \cite{cyber2021covid19}, where attackers continued their use of carefully crafted URLs to deceive users \cite{althobaiti2021idontneed, chiba2019domain}. This leads to the divulging of credentials \cite{maroofi2020areyou}, or installation of malware \cite{shibahara2017detecting}.

Such concerted use of URLs are what we define as \textit{malicious campaigns}. Regardless of the attack point, nature and time of security threats, they can be devolved down to the URL. This supplies the impetus for us to cluster and characterise seemingly distinct URLs as a motivated whole, which we do by using a practical yet effective mechanism: the hash of content pointed to the URLs themselves.

We show that we can trivially recognise suspicious collections of URLs at scale. To replicate the torrent of URLs faced by SOC teams, we gathered 311M URL submissions stretched over 2 months from VirusTotal \cite{virustotal_virustotal_2022}. This serves as a strong parallel for SOC issues, as our URLs originate from varying sources, which resembles how SOCs are alerted of URLs via multiple SIEM events. Further, our data has labels from security vendors of varying origins and quality, not unlike the diverse labelling seen in SIEM alerts. Thus, by applying content-hash grouping to this data, 2.6M suspicious clusters were exposed which led to the following contributions:

\begin{itemize}[leftmargin=*]
    \item The detection of 77,810 confirmed malicious campaigns via the reported content hashes within each submission.
    \item 6 pertinent findings relating to characteristics for 59,450 single-URL and 18,360 multi-URL malicious campaigns.
    \item 3 selected case studies confirming the impact of campaigns and the shifting tactics used by attackers to evade defences.
    
\end{itemize}

The contributions above have allowed us to quantify and articulate the influence of URL campaigns on threat intelligence efforts, which aids SOC personnel in prioritising the most serious threats and thus, amplifies their proficiency.

\section{Background \& Related Work}
\label{sec:background}

Complexity and persistence of cyber-attacks surge to new heights at scale, and threat intelligence plays an ever vital role in helping human security teams manage an otherwise overwhelming volume of information \cite{afzaliseresht_logs_2020}. It is a culmination of the old adage of ``knowing your enemy'', where the emphasis has shifted to proactive defence instead of purely reactive approaches \cite{husak_survey_2019}. We know that adversaries eschew isolated attacks in favor of coordinated campaigns \cite{li_intelligence-driven_2019}, which only underlines the gravity of the work herein.

Given that campaigns are central to threat intelligence, the research gap we strive to address is the measurement of orchestrated malicious use of URLs. Colloquially known as links, URLs are simply pointers to Internet content \cite{rfc1738}. Their simplicity and ubiquity invites abuse within multiple contexts, from social media \cite{lee2013warningbird}, to text messages \cite{reaves2016detecting}, and compromised devices \cite{zuo2017smartgen}. For SOC teams defending their systems, URLs are key Indicators of Compromise (IOC) that feed into human decisions \cite{zhu_chainsmith_2018}. For end-users, their deceptive use \cite{quinkert2020bethephisher, albakry2020what} is best exemplified in the second example below: 

\vspace{0.3em}
\begin{mdframed}[backgroundcolor=white!10,rightline=true,leftline=true,topline=true,bottomline=true,roundcorner=2mm,everyline=true,nobreak=false,innerleftmargin=0.5em,innerrightmargin=0.5em]
\small
\texttt{\textbf{Scheme \ Sub-domain \ Domain \ \ \ \ Suffix \ \ Path}}\\
        \texttt{https \ \ support \ \ \ \ apple \ \ \ \ \ com \ \ \ \ \ /en-us/}\\
        \texttt{http \ \ \ mail \ \ \ \ \ \ \ get-apple \ support \ /index.php}
\end{mdframed}

The \textbf{scheme} for most URLs remains \texttt{http}, though rising malicious use of \texttt{https} is noted in the prior art \cite{fbi2019cyber}. \textbf{Sub-domains} are fully within the purview of attackers, and is oft-used to disguise URLs \cite{quinkert2020bethephisher}. \textbf{Domain} rights are restricted depending on the \textbf{suffix} in question. For example, only Apple can register against \texttt{com.au} and \texttt{com.sg}, via resellers entrusted by their respective governing entities: auDA \cite{auda2020about} and SGNIC \cite{sgnic_singapore_2022}. \textbf{Path} refers to the exact location of a page, file or asset (i.e., the actual content) on the addressed server.

\subsection{VirusTotal \& Vendor Labelling}

Prior focus on harmful URLs has often treated them as individual pieces, with works ranging from detection at the domain name level \cite{zhauniarovich2018asurvey, hu2021assessing, desmet2021premadona, tian2018needle, szurdi2017email, miramirkhani2018panning}, deep-learning classification of URL strings \cite{das2020deep, sameen2020phishhaven, oprea2018made}, and the binaries that the URLs point to \cite{lever2017lustrum, zuo2017smartgen, li2018jsgraph}. Given the historical scale of the issue \cite{ma2011learning}, streams of suspicious URLs are sent to online security vendors, with each employing customised engines to classify malicious URLs with varying levels of success \cite{peng2019opening}.

Building atop of such services, the VirusTotal (VT) platform acts as a results aggregator \cite{song2016learning}. Essentially, it allows convenient querying of suspicious URLs across multiple vendors. While this offers simplicity, it is noted in prior works that vendor labels do not always agree, and that VT based defensive mechanisms are not sufficient \cite{sharif2018predicting, zhu_benchmarking_2020, oest_phishfarm_2019, salem_maat_2021}. As such, we carefully distill the labels within our dataset, and reinforce them with results from Google Safe Browsing (GSB) \cite{GSB}.

\begin{figure}[t]
\centering
\includegraphics[width=1.0\linewidth]{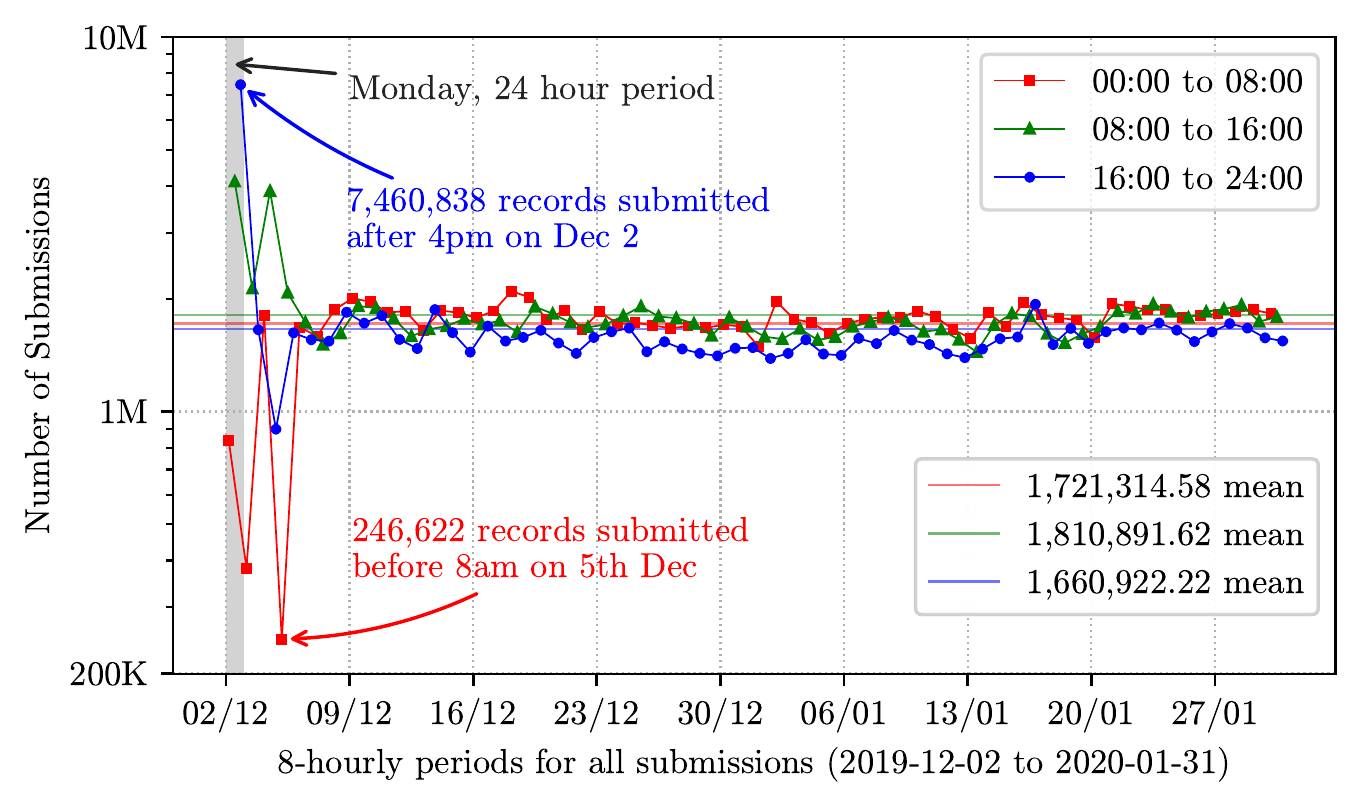}
\caption{Scan times (in UTC) of all submissions}
\label{fig:submission-scan-times}
\end{figure}

\begin{table}[b]
\centering
    \begin{tabular}{r | l}
    
    \textbf{Field} &
    \textbf{Examples}
    \\

    url
    &
    http://mail.get-apple[.]support/index.php
    \\

    content hash
    &
    f56866ad4ef7ae7c7...dc2dc840230f4215e
    \\
    
    positives
    &
    $0 \le n < N$, where $N$ is max vendors
    \\
    
    \end{tabular}

\caption{Metadata within submissions}
\label{tab:submission}
\end{table}

\section{Approach}
In this section, we detail each step of our content-hash based clustering of malicious campaigns, the secondary verification that was performed, and finally, show the potential impacts of content-hash detection.

\subsection{Dataset Description}

Our goal was to approximate the constant stream of URLs surfaced by SIEM systems. To that end, we used 311,588,127 URL submissions over a 2-month duration (2019-12-02 to 2020-01-30) gathered with kind assistance from VirusTotal. Table~\ref{tab:submission} shows selected metadata fields within each submission. Here, we focus on the ``content hash'' and ``URL'' fields, with the former being the computed hash of the \textit{content} to which the latter points. That is, if both \texttt{bad.com/mal.exe} and \texttt{new.com/run.exe} point to the same malicious executable, this would result in the \textit{same content hash}.

We note that 11.39\% of submissions (over 35.4M) had missing content hashes. These exhibited no clear patterns and were uniformly spread, thus ruling out periods of accidental corruption. Further, 77M \textit{unique} URLs were found, pointing to a naive mean rate of re-submission of 4.044 per URL. Of these, 24.5M URLs (31.9\%) were flagged by at least one vendor as malicious. Submissions were checked by an average of 71.9 vendors, with a majority (94.18\%) checked by 72.

The dataset replicates the temporal characteristics of SIEM alerts for an SOC team of a global corporation. As seen in Figure \ref{fig:submission-scan-times}, daily scan volumes were divided into three 8-hour chunks starting midnight, 8am and 4pm. With no timezone data, all times were taken to be UTC. Outliers were observed in daily volumes from December 2\textsuperscript{nd} to 5\textsuperscript{th}, but after fitting a linear model to our weekly averages, we found a slope of -0.0290, which points to a steady rate of submissions.

\subsection{Content Hash Clustering}
\label{sec:method}

\begin{figure}[t]
\centering
\includegraphics[width=1.0\linewidth]{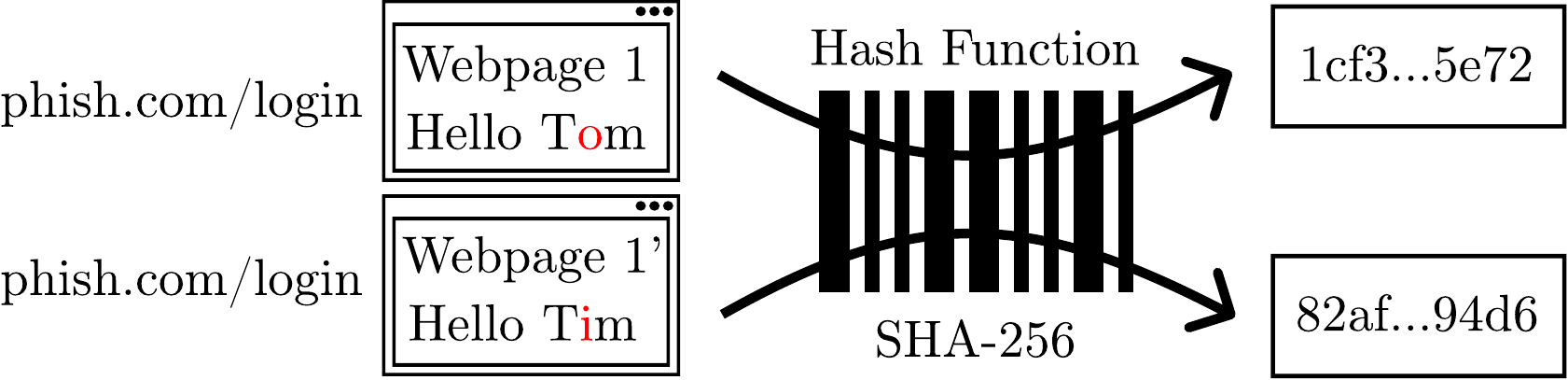}
\caption{Separate hash codes for 2 similar web-pages}
\label{fig:hash}
\end{figure}

Once submitted, the content pointed to the URL is retrieved and a SHA-256 hash function is applied. Figure \ref{fig:hash} illustrates how \texttt{Webpage 1} results in the unique identifier \texttt{1cf3...5e72}. When a single character is modified, the result changes to \texttt{82af...94d6}. This allows us to group seemingly disparate submissions: if their URLs result in identical hash codes, we can assume they point to the same content. Thus, one malicious URL would imply that all other URLs within a cluster were malicious as well. It must be noted that URLs may be part of \textit{multiple} clusters. Consider an adversary that uses the same phishing URL repeatedly but changes the target web-page over time. This means that said URL would resolve to multiple contents, and thus, multiple hash codes.

\subsection{Flagged Clusters}
\begin{figure}[t]
\centering
\includegraphics[width=1.0\linewidth]{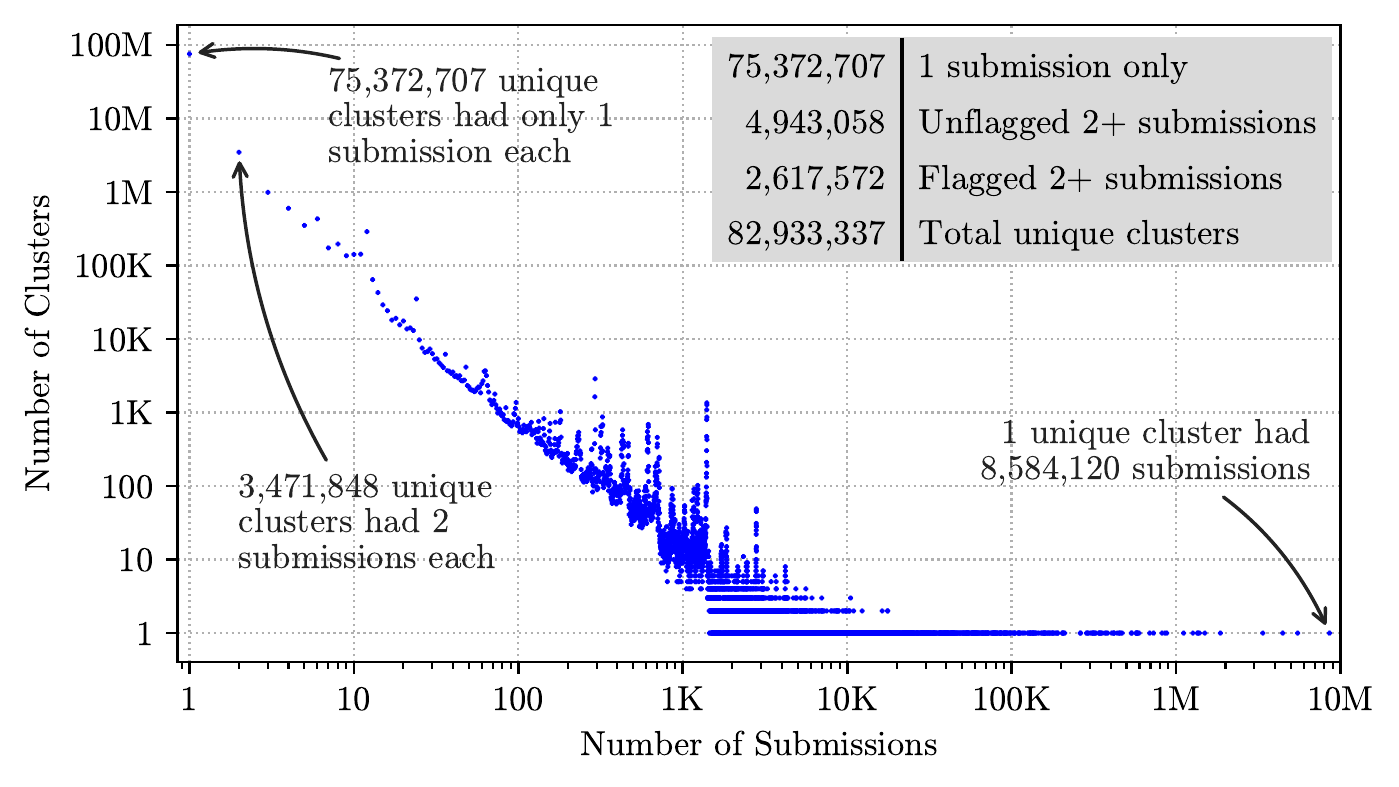}
\caption{Cluster distribution by submission counts}
\label{fig:clusters_ditribution}
\end{figure}

Accordingly, we are able to organise the 311M submissions into the 82.9M total unique clusters seen in Figure \ref{fig:clusters_ditribution}, where we see that a majority 75.3M had only one submission attached (extreme top left of figure). That is, given a particular submission $S$ with a unique content hash $H$, then $S_H=1$. These are ignored, given our focus on campaigns. We also see that one cluster had 8.5M records (extreme right) which we theorise to be a standard 404 page repeatedly submitted.

To clarify our method, we define the following:

\begin{itemize}[leftmargin=*]
    \item \textbf{Positives}: As seen in Table \ref{tab:submission}, each submission's \texttt{positives} field indicates the number of vendor flags against it.
    \item \textbf{Mean Positive Score}: This is simply the sum of all \textbf{Positives} divided by the number of submissions in each cluster.
    \item \textbf{Unflagged Clusters}: Clusters that have a mean positive score of 0.0  (i.e., no vendor flag in any of its submissions).
    \item \textbf{Flagged Clusters}: Conversely, this is when clusters have \textit{at least one} vendor hit (i.e., mean positive score > 0.0).
\end{itemize}

As seen in Figure \ref{fig:clusters_ditribution}'s table, 7.5M clusters had $S_H \ge 2$. Of those, 2.6M (34.6\%) had at least one submission flagged by a vendor, signalling the likelihood of their maliciousness.

\subsection{Secondary Verification} %

To further our analysis, we needed to confirm the malicious nature of the 2.6M flagged clusters. We cannot rely on original vendor labels as there is contention as to the correct threshold to apply. \cite{catakoglu2016automatic,hong2018you,miramirkhani2018panning,razaghpanah2018apps,sarabi2018characterizing,tian2018needle,zuo2017smartgen} have set their thresholds for maliciousness to $t=1$, with $t=2$ being used in the previous works~\cite{oprea2018made,sharif2018predicting,wang2014whowas}, and even $t=5$ in the prior work~\cite{le2018urlnet}. Further, there is an opportunity to compare existing labels against present day data. Thus, to check if clusters were indeed malicious, we implemented secondary verification via GSB, which is noted to be the largest block-list \cite{bell2020analysis}. We then expand our definition:

\begin{itemize}[leftmargin=*]
    \item \textbf{Malicious Campaigns}: Flagged clusters that are further verified by GSB as containing \textit{at least one} malicious URL.
\end{itemize}

Thus, if a cluster had submissions with positive scores of 7, 7 and 8, then its mean would be 7.33. Conversely, a cluster with a single flag amongst 10K submissions would score 0.0001. Subsequently, if either contained a URL blocked by GSB, then they are deemed as malicious campaigns.

\subsection{Initial Findings}

Using this direct approach, we discovered 77,810 confirmed malicious campaigns, which represented 2.97\% of the 2.6M aforementioned flagged clusters. These campaigns had at least one URL within them marked by GSB as either malicious (\texttt{MALWARE}), deceptive (\texttt{SOCIAL\_ENGINEERING}), or both.

\begin{finding}[With a 2.97\% detection rate, we see the limitations of current block-list approaches] On average, the mean positive score for these campaigns stands at 6.2 ($\sigma$ 3.05). Using this as a threshold, 605,323 other flagged clusters with higher mean positive scores would have eluded our secondary verification. That is, if assuming all 2.6M flagged clusters were malicious, GSB would have missed 605,323 campaigns containing 2,034,777 malicious URLs.
\end{finding}

\begin{finding}[Performance of GSB secondary verification drops as campaigns use more URLs]
We observed that detection by GSB degrades as campaigns grow larger, with a 13.27\% rate for campaigns that deploy $\geq100$ URLs as compared to 74.32\% for ones with $\leq100$ URLs. The largest campaigns exhibit even lower rates at 2.46\%, which may be attributed to the limited life-spans of malicious URLs. This agrees with the work in \cite{bell2020analysis}, which reports that fewer URLs remained in block-lists (e.g., GSB) as time progresses. This hints at the ephemeral nature of such URLs, and the fact that block-lists do not enforce a permanence policy although it was also noted that numerous URLs tend to reappear after removal. Similarly, vendors within the dataset struggle to detect URLs as campaign sizes grow. We observed that on average, six vendors detecting campaigns of size 100 and below, and only four detecting larger campaigns (size over 100).

\end{finding}

\subsection{Broader Impact of Content Hashes}
A consequence of our work is perhaps that the detection of threats within SIEM alerts can be improved if SOC teams can factor in content-hashes at the triage stage. Blocking singular URLs works on past but not \textit{future} threats. Using content-hashes, previously unseen threats could be rapidly identified in the absence of other obvious malicious connections.

We see evidence of this with the small malicious campaign below, where one URL (everyday-vouchers.com) remained unflagged for six submissions before being flagged by GSB on 2020-01-02. Soon after, another URL (sweepstakehunter.com) bearing the \textit{same content hash} was observed as remaining unflagged for 5 submissions to the end of the collected data.

\vspace{0.3em}

\begin{mdframed}[backgroundcolor=white!10,rightline=true,leftline=true,topline=true,bottomline=true,roundcorner=2mm,everyline=true,nobreak=false,innerleftmargin=0.5em,innerrightmargin=0.5em] 
    \small
    \textbf{Malicious Campaign ID}: 16f3-21a6 \\
    \textbf{Mean Positive Score}: 0.08333 (across 12 submissions)\\
    \textbf{URLs}: everyday-vouchers[.]com, sweepstakehunter[.]com
\end{mdframed}

One year later, \textit{both} URLs were marked as malicious by GSB in our secondary checks. In this case, a SIEM threat analyser based on content-hash would have raised the priority alert for the second URL as soon as it had appeared. It is noted that attackers could automate the mutation of malicious content for each end-user visit. This would evade content-hash detection as presumably, each URL results in a unique hash. Accordingly, teams can fall back on methods such as linking IP addresses and server signatures, or on HTML comparison methods such as in \cite{li2019stacking}. Nonetheless, content hashes are still useful for clustering URL alerts before more advanced mechanisms are brought to bear.

\section{Measuring Malicious Campaigns}
\label{sec:malicious}

\begin{table}[t]
\centering
  
 \begin{tabular}{%
    r|%
    l%
 }

    59,450 &
    Campaigns with a single unique URL \\
    18,360 &
    Campaigns with multiple unique URLs \\
    \textbf{77,810} &
    \textbf{Total Campaigns} \\[0.4em]

    59,450 &
    URLs in Single-URL Campaigns \\
    9,900,146 &
    URLs in Multi-URL Campaigns \\
    \textbf{9,959,596} &
    \textbf{Total Unique URLs} \\[0.4em]

    276,801 &
    Submissions in Single-URL Campaigns \\
    37,880,096 &
    Submissions in Multi-URL Campaigns \\
    \textbf{38,156,897} &
    \textbf{Total Submissions} \\[0.4em]

\end{tabular}

\caption{Characteristics of found malicious campaigns}
\label{tab:campaigns}
\end{table}

Preceding sections may have only affirmed an instinctual knowledge within SOC teams: that the daily torrent of SIEM alerts about suspicious URLs are rarely isolated events. We thus dive deeper into our discovered malicious campaigns.

\begin{finding}[Vast majority of malicious URLs are from campaigns employing multiple URLs]
Table \ref{tab:campaigns} shows the general details of confirmed campaigns, which we have broadly designated into campaigns that have either a single unique URL, or multiple unique URLs. A majority (76.4\%) of the found campaigns consists of single URLs (i.e., the same URL is repeatedly submitted) but these only contributed 59,450 (0.6\%) unique URLs and 276,801 (0.73\%) submissions. The 18,360 multi-URL campaigns provided the bulk of URLs and submissions, lending credence to our hypothesis that masses of URLs should be treated as coherent campaigns.
\end{finding}

\begin{figure}[b]
\centering
\includegraphics[width=1.0\linewidth]{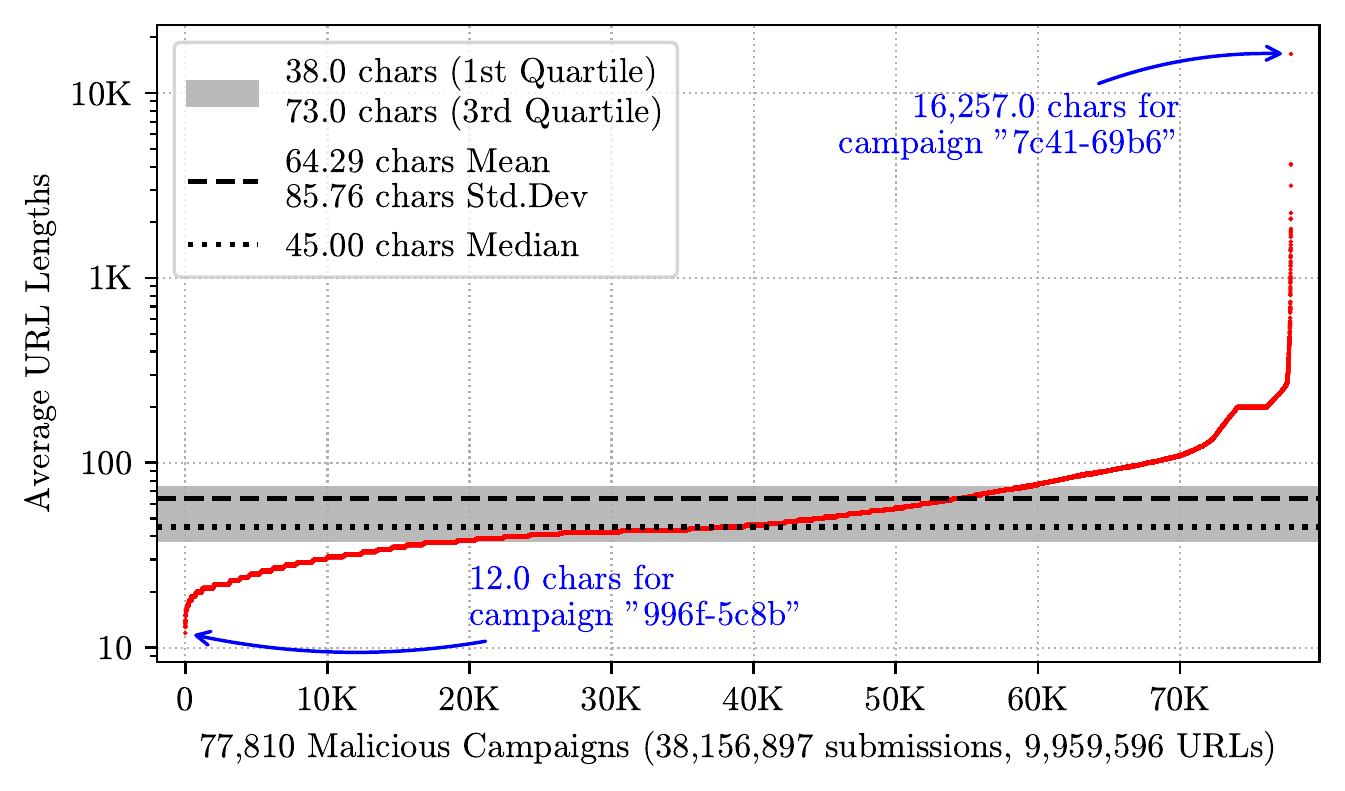}
\caption{Mean URL lengths of 77,810 campaigns}
\label{fig:url-lengths-campaigns}
\end{figure}

\begin{finding}[Campaigns favor the use of longer URLs]
The average URL lengths across campaigns (i.e., mean of means) stands at 64.29 characters. There is a standard deviation of 85.76 indicating a high degree of variance which we see in Figure \ref{fig:url-lengths-campaigns} with some campaigns on the extreme right employing URLs with tens of thousands of characters. 50.89\% of campaigns have average URL lengths of between 38.0 (1st quartile) and 73.0 (3rd quartile). This is likely due to attackers relying on increasingly longer URLs to evade detection. As awareness increases in end-users, they grow to distrust shorter URLs pointing to unknown locations \cite{albakry2020what} or short URLs with added terms in their domain names \cite{quinkert2020bethephisher}.
\end{finding}

\subsection{Campaign Metrics}

Next, to systematically quantify malicious campaign behaviours, we define and calculate the following metrics:

\begin{itemize}[leftmargin=*]
    \item \textbf{Campaign size}: Total number of URLs deployed.

    \item \textbf{Source distribution}: Ratio of the campaign size to the number of submissions. A 100\% source distribution indicates that a different URL is used for every submission. This quantifies the rate at which attackers attempt different URLs, and the URL diversity within a campaign.
    
    \item \textbf{Campaign footprint}: Here, we consider metrics from both attackers (via a count of unique URLs in a campaign), and victims (number of times URLs were reported). The following expression is used to compute the footprint: $fp = \mu U + (1-\mu)S$, where $U$ represents the count of URLs in a campaign and $S$ denotes the number of times these URLs were submitted. $\mu$ represents the weight assigned to each metric, with equal weights of 0.5 set for both sides and $\mu$ able to prioritize certain metrics.
    
    \item \textbf{Domain diversity}: Ratio of unique domains to the number of URLs in a campaign, where a 100\% score indicates that each URL has used a unique domain. This metric aids us to understand if a campaign is targeting specific global brands, or if it distributes attacks across a range of users. Conversely, low domain diversity can suggest a highly targeted campaign towards an organization.

    \item \textbf{Sub-domain diversity}: Same as domain diversity above but for sub-domains within the URLs of a campaign.

    \item \textbf{Detection}: Here, we evaluate detection rates using two metrics: GSB and the original vendors within the data. A 100\% GSB detection means that all of a campaign's URLs were flagged by GSB. For vendors, we report the mean number of vendors who flagged a campaign's URLs. This helps us understand the general detection rate of URLs.
\end{itemize}

\subsection{Campaign Insights}
\label{ssec:campaign-insights}

\begin{finding}[Most campaigns deploy different URLs in their attacks]
The average source distribution for all campaigns stands at 61.58\%. We observed that 6,160 campaigns with an average campaign size of 146 URLs (footprint of 160) had source distribution of 98.75\%. Moreover, 5,498 campaigns had source distribution of 100\%. These high source distributions show that campaigns are frequently creating new attack URLs, and also signals that traditional allow/block-lists will not effectively detect URLs in many campaigns.  

We also observed that for 1,040 larger campaigns with campaign size greater than 100 URLs, the source distribution reduced to 35.95\%. For such campaigns, the average campaign size, footprint, and number of reported submissions were 5,132, 19,682, and 34,233, respectively. Moreover, we observed that the largest campaign with a campaign size of 1,391,080 URLs (footprint at 4,987,600, and submissions at 8,584,120) had a source distribution of 16.21\%.

These findings indicate that the source distribution of a campaign decreases with increases in campaign size. This may be attributed to the fact that as the number of URLs increases, the corresponding reports from users (i.e., submissions) increases non-linearly. This indicates that larger campaigns may be using domain generation algorithms (DGAs) to generate a swathe of domain names for both targeting, and command and control. This large number makes it arduous for vendors to take down URLs from such campaigns, as infected computers will attempt to contact some of these domain names daily to receive updates or commands. 
\end{finding}

\begin{finding}[We discovered several campaigns impersonating well-known global brands] Given the finding above, we found campaigns using multitudes of URLs to mislead victims by including company or product names. The example below shows a campaign that targeted Apple with 4,081 unique URLs. It used a combination of 9 sub-domains, 12 domains, and 7 suffixes, with the ratio of unique domains to URLs standing at 0.29\%, using the following variations:

\vspace{0.3em}
\begin{mdframed}[backgroundcolor=white!10,rightline=true,leftline=true,topline=true,bottomline=true,roundcorner=2mm,everyline=true,nobreak=false,innerleftmargin=0.5em,innerrightmargin=0.5em]
    \small
        \textbf{Sub-Domains}: \texttt{www.apple.com}, \texttt{www.icloud.com}, \texttt{www.apple}, \texttt{icloud.com},  \texttt{apple}, \texttt{apple.com}, \texttt{icloud}, \texttt{www}, \texttt{mail}.\newline 
        \textbf{Domains}: \texttt{lcloud-com},  \texttt{online-support}, \texttt{findmy}, \texttt{get-apple}, \texttt{com-support}, \texttt{map-apple},  \texttt{id-info}, \texttt{wvvw-icloud}, \texttt{sign-in},\newline
        \texttt{map-log}, \texttt{com-fml}, \texttt{viewlocation-icloud}.\newline
        \textbf{Top-Level Domains}: \texttt{us}, \texttt{in}, \texttt{support}, \texttt{live}, \texttt{review}, \texttt{com}, \texttt{mobi}
\end{mdframed}

We see the precise targeting of not only the company name, but its various associated products. The particular content hash for this campaign was submitted 104,311 times, and we observed that an average 11 vendors flagged URLs within this campaign. This compares well against the average 3.4973 flags across all 311M submissions. Surprisingly, we also observed that GSB only detected two of the URLs.
\end{finding}

\begin{finding}[Campaigns had an average domain and sub-domain diversity of 66.57\% and 21.45\%, respectively, with specific targeted organizations]
We observed higher diversity in domain names compared to that of sub-domains. This is because attackers are limited to certain guidelines when registering domain names whereas sub-domains can be chosen freely. Due to this flexibility, campaigns targeting particular organizations can use sub-domain names specific to the target organization, such as in the targeted phishing above. All in all, we observed that 153 campaigns had domain and sub-domain diversities of 2.29\% and 1.61\%, respectively. This indicates that attackers used fixed sets of domains and sub-domains targeting a few victim brands.
\end{finding}

\begin{finding}[98.93\% (1,572/1,589) of URLs within a known malware campaign used transport layer protocol (TLS)] Continuing the trend of diverse malicious URLs, we found a campaign aimed at installing malware into victim machines, with a size of 1,589 URLs and a source distribution of 16.57\%. The campaign used 261 unique domain names and a total of 1,589 URLs which were submitted 9,589 times. 1,175 of the URLs had the domain \texttt{ubar-}\texttt{pro4.ru}, which was used specifically for malware distribution \cite{jeosandbox}. This finding is in line with \cite{sophos}, which reports that up to 25\% of malware rely on TLS. Encryption is a potent evasive tactic that obfuscates code and/or the path component of URLs, and we note that these URLs lure users to download malware with file extensions such as: \texttt{exe} (19), \texttt{js} (28), \texttt{zip} (7), and \texttt{rar} (3).

We also discovered that the campaign's URLs were embedded with pointers to malicious torrent files, which allows malware to be downloaded via numerous torrent peers rather than a single server. This enables strong persistence of malicious content even when trackers are unavailable. We observed 46 (2.89\%) URLs with such capability, with very low detection rates for GSB and VT vendors. From the 1,589 unique URLs, only two were detected by GSB and on average, only one vendor flagged the URLs within.
\end{finding}

\subsection{Fileless Malware}
\label{ssec: case study fileless malware}

Campaigns also exhibit sophisticated tactics that leverage native pre-installed system tools (colloquially named ``living off the land''). Such fileless malware attacks forego the need for payloads that can trigger existing disk scanning tools.

\begin{finding}[We observed 23 campaigns that leverage a system's pre-installed software for compromise]
The URLs within such campaigns are representative of a three-stage process: First, the infection typically starts with phishing emails rigged with deceptive URLs. Second, a malicious payload is downloaded using a command-line tool (e.g., \textit{cmd.exe}). The Windows PowerShell tool can also be invoked with heavily obfuscated code to evade analysis. The payload is kept in-memory rather than on-disk, where it is finally executed, whereupon its plugins are loaded and a communication loop is entered that fetches tasks from command servers. The example below shows malicious code embedded within a URL that downloads and executes \texttt{bad.exe} via PowerShell:

\vspace{0.3em}

\begin{mdframed}[backgroundcolor=white!10,rightline=true,leftline=true,topline=true,bottomline=true,roundcorner=2mm,everyline=true,nobreak=false,innerleftmargin=0.5em,innerrightmargin=0.5em]
    \small
   \sloppy
    \url{http://xxx.xxx.xxx[.]xxx/public/invokefunction\&function= call\_user\_func\_array\&vars[0]=system\&vars[1][]=}\textbf{cmd.exe}$\backslash$/c$\backslash$\textbf{powershell}$\backslash$(new-object$\backslash$System.Net.WebClient). \textbf{DownloadFile}\newline
    ('www.bad.com/download.exe','$\backslash$SystemRoot$\backslash$ /Temp/\textbf{bad.exe}');\textbf{start}$\backslash$SystemRoot$ \backslash$/Temp/\textbf{bad.exe}
\end{mdframed}

\vspace{0.3em}

Upon further analysis, we found other legitimate system tools abused for malicious tasks, including:

\vspace{0.3em}
\begin{mdframed}[backgroundcolor=white!10,rightline=true,leftline=true,topline=true,bottomline=true,roundcorner=2mm,everyline=true,nobreak=false,innerleftmargin=0.5em,innerrightmargin=0.5em] 
    \small
    wmiprvse.exe, svchost.exe, conhost.exe, lsm.exe, sysstem.exe, lsass.exe, winlogon.exe, spoolsv.exe, csrss.exe, securityhealthservice.exe, services.exe, wininit.exe, cmd.exe, explorer.exe, smss.exe, wudfhost.exe, yourphone.exe, dwm.exe, taskhost.exe, csrss.exe, system.exe, regasm.exe, idle.exe
\end{mdframed}

\end{finding}

\section{Conclusion}
\label{sec:conclusion}

We hope that the \total{finding} findings presented can aid SOC personnel rapidly triage and act upon the most pertinent threats. Traditional approaches must be improved as we cannot rely on the wholesale lumping of URLs into binary benign or malicious classifications. For deep-learning based detectors, threat intelligence such as this could help tease out features and connections amongst disparate URLs. The already bewildering array of tricks employed by attackers are ever evolving, as seen in the range of domains, sub-domains and suffixes used to deceive both defences and end-users in Section \ref{ssec:campaign-insights}. We also note the widely variant URL lengths and propensity for longer URLs as seen in Section \ref{sec:malicious}, and the increased sophistication with fileless malware in Section \ref{ssec: case study fileless malware}.

However, such malicious attacks ultimately hinges on a singular pivot point: the humble URL. This is true regardless of the point wherein the URL was introduced into the SIEM, be it within a suspicious email to an employee, or as an outgoing suspicious request from a possibly compromised corporate device. Accordingly, measuring suspicious URLs as part of coherent campaigns allows for more refined, robust and timely detection protocols within SOCs. Perhaps a key takeaway is that we can already achieve much better threat intelligence by simply inspecting content hashes, without necessarily and immediately leaping to deep-learning or other more advanced heuristic approaches.

Thus in pursuit of the above goal, the work herein is a culmination of extensive research and analysis into 311M records submitted to VirusTotal a 2-month period ending Jan 2020. While there exist extensive studies on malicious URLs, there remains a gap in our understanding of them as concerted \textit{campaigns}. Our work clusters URLs via their metadata and doubly verifies them as malicious before further analysis. This offers practical insights into the nature of campaigns along dimensions such as individual URL attributes and targeted victim brands. We also explored case studies that illustrate various adversarial techniques that can circumvent defences. Future work includes looking at measures to further group campaigns together. That is, to see if there are avenues for us to connect disparate campaigns into a meta-campaign attributable to a common attacker.

\section*{Acknowledgements}
The work has been supported by the Cyber Security Research Centre Limited whose activities are partially funded by the Australian Government’s Cooperative Research Centres Programme $\bullet$ This work was supported by resources provided by the Pawsey Supercomputing Centre with funding from the Australian Government and the Government of Western Australia $\bullet$ This research was supported by use of the Nectar Research Cloud and by the Tasmanian Partnership for Advanced Computing (TPAC).  The Nectar Research Cloud is a collaborative Australian research platform supported by the NCRIS-funded Australian Research Data Commons (ARDC).

\bibliographystyle{unsrt}
\bibliography{references}

\begin{thebibliography}{10}

\bibitem{kokulu_matched_2019}
Faris~Bugra Kokulu, Ananta Soneji, Tiffany Bao, Yan Shoshitaishvili, Ziming
  Zhao, Adam Doupé, and Gail-Joon Ahn.
\newblock Matched and {Mismatched} {SOCs}: {A} {Qualitative} {Study} on
  {Security} {Operations} {Center} {Issues}.
\newblock In {\em Proceedings of the 2019 {ACM} {SIGSAC} {Conference} on
  {Computer} and {Communications} {Security}}, {CCS} '19, pages 1955--1970, New
  York, NY, USA, November 2019. Association for Computing Machinery.

\bibitem{islam_multi-vocal_2019}
Chadni Islam, Muhammad~Ali Babar, and Surya Nepal.
\newblock A {Multi}-{Vocal} {Review} of {Security} {Orchestration}.
\newblock {\em ACM Computing Surveys}, 52(2):37:1--37:45, April 2019.

\bibitem{rosso_saibersoc_2020}
Martin Rosso, Michele Campobasso, Ganduulga Gankhuyag, and Luca Allodi.
\newblock {SAIBERSOC}: {Synthetic} {Attack} {Injection} to {Benchmark} and
  {Evaluate} the {Performance} of {Security} {Operation} {Centers}.
\newblock In {\em Annual {Computer} {Security} {Applications} {Conference}},
  {ACSAC} '20, pages 141--153, New York, NY, USA, December 2020. Association
  for Computing Machinery.

\bibitem{akinrolabu_challenge_2018}
Olusola Akinrolabu, Ioannis Agrafiotis, and Arnau Erola.
\newblock The challenge of detecting sophisticated attacks: {Insights} from
  {SOC} {Analysts}.
\newblock In {\em Proceedings of the 13th {International} {Conference} on
  {Availability}, {Reliability} and {Security}}, {ARES} 2018, pages 1--9, New
  York, NY, USA, August 2018. Association for Computing Machinery.

\bibitem{sundaramurthy_tale_2014}
Sathya~Chandran Sundaramurthy, Jacob Case, Tony Truong, Loai Zomlot, and Marcel
  Hoffmann.
\newblock A {Tale} of {Three} {Security} {Operation} {Centers}.
\newblock In {\em Proceedings of the 2014 {ACM} {Workshop} on {Security}
  {Information} {Workers}}, {SIW} '14, pages 43--50, New York, NY, USA,
  November 2014. Association for Computing Machinery.

\bibitem{fireeye_what_2022}
FireEye.
\newblock What is {SIEM} and how does it work?, 2022.

\bibitem{cofense_siem_2022}
Cofense.
\newblock {SIEM} {Alerts} {\textbar} {What} is a {SIEM} {Alert} and {How}
  {Does} {It} {Work}?, 2022.

\bibitem{najafi_malrank_2019}
Pejman Najafi, Alexander Mühle, Wenzel Pünter, Feng Cheng, and Christoph
  Meinel.
\newblock {MalRank}: a measure of maliciousness in {SIEM}-based knowledge
  graphs.
\newblock In {\em Proceedings of the 35th {Annual} {Computer} {Security}
  {Applications} {Conference}}, {ACSAC} '19, pages 417--429, New York, NY, USA,
  December 2019. Association for Computing Machinery.

\bibitem{shu_threat_2018}
Xiaokui Shu, Frederico Araujo, Douglas~L. Schales, Marc~Ph. Stoecklin, Jiyong
  Jang, Heqing Huang, and Josyula~R. Rao.
\newblock Threat {Intelligence} {Computing}.
\newblock In {\em Proceedings of the 2018 {ACM} {SIGSAC} {Conference} on
  {Computer} and {Communications} {Security}}, {CCS} '18, pages 1883--1898, New
  York, NY, USA, October 2018. Association for Computing Machinery.

\bibitem{ban_combat_2021}
Tao Ban, Ndichu Samuel, Takeshi Takahashi, and Daisuke Inoue.
\newblock Combat {Security} {Alert} {Fatigue} with {AI}-{Assisted}
  {Techniques}.
\newblock In {\em Cyber {Security} {Experimentation} and {Test} {Workshop}},
  {CSET} '21, pages 9--16, New York, NY, USA, August 2021. Association for
  Computing Machinery.

\bibitem{oprea2018made}
Alina Oprea, Zhou Li, Robin Norris, and Kevin Bowers.
\newblock Made: Security analytics for enterprise threat detection.
\newblock In {\em Proceedings of the 34th Annual Computer Security Applications
  Conference}, pages 124--136, 2018.

\bibitem{merah_ontology-based_2021}
Yazid Merah and Tayeb Kenaza.
\newblock Ontology-based {Cyber} {Risk} {Monitoring} {Using} {Cyber} {Threat}
  {Intelligence}.
\newblock In {\em The 16th {International} {Conference} on {Availability},
  {Reliability} and {Security}}, {ARES} 2021, pages 1--8, New York, NY, USA,
  August 2021. Association for Computing Machinery.

\bibitem{ahmed_survey_2016}
Mohiuddin Ahmed, Abdun Naser~Mahmood, and Jiankun Hu.
\newblock A survey of network anomaly detection techniques.
\newblock {\em Journal of Network and Computer Applications}, 60:19--31,
  January 2016.

\bibitem{cyber2021covid19}
{Australian Signals Directorate’s Australian Cyber Security Centre}.
\newblock {Threat update: COVID-19 malicious cyber activity 20 April 2020}.
\newblock
  \url{https://www.cyber.gov.au/sites/default/files/2020-04/ACSC-Threat-Update-COVID-19-Malicious-Cyber-Activity-20200420.pdf}.
\newblock Accessed 2021-05-12.

\bibitem{althobaiti2021idontneed}
Kholoud Althobaiti, Nicole Meng, and Kami Vaniea.
\newblock I don’t need an expert! making url phishing features human
  comprehensible.
\newblock In {\em In CHI Conference on Human Factors in Computing Systems (CHI
  ’21)}, May 2021.

\bibitem{chiba2019domain}
Daiki Chiba, Ayako~Akiyama Hasegawa, Takashi Koide, Yuta Sawabe, Shigeki Goto,
  and Mitsuaki Akiyama.
\newblock Domainscouter: Understanding the risks of deceptive idns.
\newblock In {\em 22nd International Symposium on Research in Attacks,
  Intrusions and Defenses ({RAID} 2019)}, pages 413--426, Chaoyang District,
  Beijing, September 2019. {USENIX} Association.

\bibitem{maroofi2020areyou}
Sourena Maroofi, Maciej Korczy\'{n}ski, and Andrzej Duda.
\newblock Are you human? resilience of phishing detection to evasion techniques
  based on human verification.
\newblock In {\em Proceedings of the ACM Internet Measurement Conference}, IMC
  '20, page 78–86, New York, NY, USA, 2020. Association for Computing
  Machinery.

\bibitem{shibahara2017detecting}
Toshiki Shibahara, Yuta Takata, Mitsuaki Akiyama, Takeshi Yagi, and Takeshi
  Yada.
\newblock Detecting malicious websites by integrating malicious, benign, and
  compromised redirection subgraph similarities.
\newblock In {\em 2017 IEEE 41st Annual Computer Software and Applications
  Conference (COMPSAC)}, volume~1, pages 655--664, 2017.

\bibitem{virustotal_virustotal_2022}
VirusTotal.
\newblock {VirusTotal}: {How} it works, 2022.

\bibitem{afzaliseresht_logs_2020}
Neda Afzaliseresht, Yuan Miao, Sandra Michalska, Qing Liu, and Hua Wang.
\newblock From logs to {Stories}: {Human}-{Centred} {Data} {Mining} for {Cyber}
  {Threat} {Intelligence}.
\newblock {\em IEEE Access}, 8:19089--19099, 2020.
\newblock Conference Name: IEEE Access.

\bibitem{husak_survey_2019}
Martin Husák, Jana Komárková, Elias Bou-Harb, and Pavel Čeleda.
\newblock Survey of {Attack} {Projection}, {Prediction}, and {Forecasting} in
  {Cyber} {Security}.
\newblock {\em IEEE Communications Surveys Tutorials}, 21(1):640--660, 2019.
\newblock Conference Name: IEEE Communications Surveys Tutorials.

\bibitem{li_intelligence-driven_2019}
Yuqing Li, Wenkuan Dai, Jie Bai, Xiaoying Gan, Jingchao Wang, and Xinbing Wang.
\newblock An {Intelligence}-{Driven} {Security}-{Aware} {Defense} {Mechanism}
  for {Advanced} {Persistent} {Threats}.
\newblock {\em IEEE Transactions on Information Forensics and Security},
  14(3):646--661, March 2019.
\newblock Conference Name: IEEE Transactions on Information Forensics and
  Security.

\bibitem{rfc1738}
Tim Berners-Lee, Larry~M Masinter, and Mark~P. McCahill.
\newblock {Uniform Resource Locators (URL)}.
\newblock RFC 1738, December 1994.

\bibitem{lee2013warningbird}
Sangho Lee and Jong Kim.
\newblock {WarningBird}: {A} {Near} {Real}-{Time} {Detection} {System} for
  {Suspicious} {URLs} in {Twitter} {Stream}.
\newblock {\em IEEE Transactions on Dependable and Secure Computing},
  10(3):183--195, May 2013.
\newblock Conference Name: IEEE Transactions on Dependable and Secure
  Computing.

\bibitem{reaves2016detecting}
Bradley Reaves, Logan Blue, Dave Tian, Patrick Traynor, and Kevin~R.B. Butler.
\newblock Detecting {SMS} {Spam} in the {Age} of {Legitimate} {Bulk}
  {Messaging}.
\newblock In {\em Proceedings of the 9th {ACM} {Conference} on {Security} \&
  {Privacy} in {Wireless} and {Mobile} {Networks}}, {WiSec} '16, pages
  165--170, New York, NY, USA, July 2016. Association for Computing Machinery.

\bibitem{zuo2017smartgen}
Chaoshun Zuo and Zhiqiang Lin.
\newblock Smartgen: Exposing server urls of mobile apps with selective symbolic
  execution.
\newblock In {\em Proceedings of the 26th International Conference on World
  Wide Web}, pages 867--876, 2017.

\bibitem{zhu_chainsmith_2018}
Ziyun Zhu and Tudor Dumitras.
\newblock {ChainSmith}: {Automatically} {Learning} the {Semantics} of
  {Malicious} {Campaigns} by {Mining} {Threat} {Intelligence} {Reports}.
\newblock In {\em 2018 {IEEE} {European} {Symposium} on {Security} and
  {Privacy} ({EuroS} {P})}, pages 458--472, April 2018.

\bibitem{quinkert2020bethephisher}
Florian Quinkert, Martin Degeling, Jim Blythe, and Thorsten Holz.
\newblock Be the {Phisher} -- {Understanding} {Users}' {Perception} of
  {Malicious} {Domains}.
\newblock In {\em Proceedings of the 15th {ACM} {Asia} {Conference} on
  {Computer} and {Communications} {Security}}, {ASIA} {CCS} '20, pages
  263--276, New York, NY, USA, October 2020. Association for Computing
  Machinery.

\bibitem{albakry2020what}
Sara Albakry, Kami Vaniea, and Maria~K. Wolters.
\newblock What is this {URL}'s {Destination}? {Empirical} {Evaluation} of
  {Users}' {URL} {Reading}.
\newblock In {\em Proceedings of the 2020 {CHI} {Conference} on {Human}
  {Factors} in {Computing} {Systems}}, {CHI} '20, pages 1--12, New York, NY,
  USA, April 2020. Association for Computing Machinery.

\bibitem{fbi2019cyber}
{Federal Bureau of Investigation}.
\newblock {Cyber Actors Exploit 'Secure' Websites In Phishing Campaigns}.
\newblock \url{https://www.ic3.gov/Media/Y2019/PSA190610}.
\newblock Accessed 2021-05-24.

\bibitem{auda2020about}
{.au Domain Administration Ltd}.
\newblock {About .au Domain Administration}.
\newblock
  \url{https://www.auda.org.au/about-auda/about-au-domain-administration}.
\newblock Accessed 2021-05-24.

\bibitem{sgnic_singapore_2022}
SGNIC.
\newblock Singapore {Network} {Information} {Center} - {List} of {Registrars}.
\newblock https://www.sgnic.sg/domain-registration/list-of-registrars, 2022.

\bibitem{zhauniarovich2018asurvey}
Yury Zhauniarovich, Issa Khalil, Ting Yu, and Marc Dacier.
\newblock A survey on malicious domains detection through dns data analysis.
\newblock {\em ACM Comput. Surv.}, 51(4), July 2018.

\bibitem{hu2021assessing}
Hang Hu, Steve~T.K. Jan, Yang Wang, and Gang Wang.
\newblock Assessing browser-level defense against idn-based phishing.
\newblock In {\em 30th {USENIX} Security Symposium ({USENIX} Security 21)}.
  {USENIX} Association, August 2021.

\bibitem{desmet2021premadona}
Lieven Desmet, Jan Spooren, Thomas Vissers, Peter Janssen, and Wouter Joosen.
\newblock Premadoma: An operational solution to prevent malicious domain name
  registrations in the .eu tld.
\newblock {\em Digital Threats: Research and Practice}, 2(1), January 2021.

\bibitem{tian2018needle}
Ke~Tian, Steve~TK Jan, Hang Hu, Danfeng Yao, and Gang Wang.
\newblock Needle in a haystack: Tracking down elite phishing domains in the
  wild.
\newblock In {\em Proceedings of the Internet Measurement Conference 2018},
  pages 429--442, 2018.

\bibitem{szurdi2017email}
Janos Szurdi and Nicolas Christin.
\newblock Email typosquatting.
\newblock In {\em IMC}, pages 419--431, 2017.

\bibitem{miramirkhani2018panning}
Najmeh Miramirkhani, Timothy Barron, Michael Ferdman, and Nick Nikiforakis.
\newblock Panning for gold. com: Understanding the dynamics of domain
  dropcatching.
\newblock In {\em Proceedings of the 2018 World Wide Web Conference}, pages
  257--266, 2018.

\bibitem{das2020deep}
Arijit Das, Ankita Das, Anisha Datta, Shukrity Si, and Subhas Barman.
\newblock Deep approaches on malicious url classification.
\newblock In {\em 2020 11th International Conference on Computing,
  Communication and Networking Technologies (ICCCNT)}, pages 1--6, 2020.

\bibitem{sameen2020phishhaven}
Maria Sameen, Kyunghyun Han, and Seong~Oun Hwang.
\newblock Phishhaven—an efficient real-time ai phishing urls detection
  system.
\newblock {\em IEEE Access}, 8:83425--83443, 2020.

\bibitem{lever2017lustrum}
Chaz Lever, Platon Kotzias, Davide Balzarotti, Juan Caballero, and Manos
  Antonakakis.
\newblock A lustrum of malware network communication: Evolution and insights.
\newblock In {\em 2017 IEEE Symposium on Security and Privacy (SP)}, pages
  788--804. IEEE, 2017.

\bibitem{li2018jsgraph}
Bo~Li, Phani Vadrevu, Kyu~Hyung Lee, Roberto Perdisci, Jienan Liu, Babak
  Rahbarinia, Kang Li, and Manos Antonakakis.
\newblock Jsgraph: Enabling reconstruction of web attacks via efficient
  tracking of live in-browser javascript executions.
\newblock In {\em NDSS}, 2018.

\bibitem{ma2011learning}
Justin Ma, Lawrence~K. Saul, Stefan Savage, and Geoffrey~M. Voelker.
\newblock Learning to detect malicious {URLs}.
\newblock {\em ACM Transactions on Intelligent Systems and Technology},
  2(3):30:1--30:24, May 2011.

\bibitem{peng2019opening}
Peng Peng, Limin Yang, Linhai Song, and Gang Wang.
\newblock Opening the {Blackbox} of {VirusTotal}: {Analyzing} {Online}
  {Phishing} {Scan} {Engines}.
\newblock In {\em Proceedings of the {Internet} {Measurement} {Conference}},
  {IMC} '19, pages 478--485, New York, NY, USA, October 2019. Association for
  Computing Machinery.

\bibitem{song2016learning}
Linhai Song, Heqing Huang, Wu~Zhou, Wenfei Wu, and Yiying Zhang.
\newblock Learning from {Big} {Malwares}.
\newblock In {\em Proceedings of the 7th {ACM} {SIGOPS} {Asia}-{Pacific}
  {Workshop} on {Systems}}, {APSys} '16, pages 1--8, New York, NY, USA, August
  2016. Association for Computing Machinery.

\bibitem{sharif2018predicting}
Mahmood Sharif, Jumpei Urakawa, Nicolas Christin, Ayumu Kubota, and Akira
  Yamada.
\newblock Predicting impending exposure to malicious content from user
  behavior.
\newblock In {\em Proceedings of the 2018 ACM SIGSAC Conference on Computer and
  Communications Security}, pages 1487--1501, 2018.

\bibitem{zhu_benchmarking_2020}
Shuofei Zhu, Ziyi Zhang, Limin Yang, Linhai Song, and Gang Wang.
\newblock Benchmarking {Label} {Dynamics} of {VirusTotal} {Engines}.
\newblock In {\em Proceedings of the 2020 {ACM} {SIGSAC} {Conference} on
  {Computer} and {Communications} {Security}}, pages 2081--2083. Association
  for Computing Machinery, New York, NY, USA, October 2020.

\bibitem{oest_phishfarm_2019}
Adam Oest, Yeganeh Safaei, Adam Doupe, Gail-Joon Ahn, Brad Wardman, and Kevin
  Tyers.
\newblock {PhishFarm}: {A} {Scalable} {Framework} for {Measuring} the
  {Effectiveness} of {Evasion} {Techniques} against {Browser} {Phishing}
  {Blacklists}.
\newblock In {\em 2019 {IEEE} {Symposium} on {Security} and {Privacy} ({SP})},
  pages 1344--1361, San Francisco, CA, USA, May 2019. IEEE.

\bibitem{salem_maat_2021}
Aleieldin Salem, Sebastian Banescu, and Alexander Pretschner.
\newblock Maat: {Automatically} {Analyzing} {VirusTotal} for {Accurate}
  {Labeling} and {Effective} {Malware} {Detection}.
\newblock {\em ACM Transactions on Privacy and Security}, 24(4):25:1--25:35,
  July 2021.

\bibitem{GSB}
{Google}.
\newblock Google safe browsing.
\newblock \url{https://safebrowsing.google.com/}.
\newblock Accessed: 2021-05-24.

\bibitem{catakoglu2016automatic}
Onur Catakoglu, Marco Balduzzi, and Davide Balzarotti.
\newblock Automatic extraction of indicators of compromise for web
  applications.
\newblock In {\em Proceedings of the 25th international conference on world
  wide web}, pages 333--343, 2016.

\bibitem{hong2018you}
Geng Hong, Zhemin Yang, Sen Yang, Lei Zhang, Yuhong Nan, Zhibo Zhang, Min Yang,
  Yuan Zhang, Zhiyun Qian, and Haixin Duan.
\newblock How you get shot in the back: A systematical study about
  cryptojacking in the real world.
\newblock In {\em Proceedings of the 2018 ACM SIGSAC Conference on Computer and
  Communications Security}, pages 1701--1713, 2018.

\bibitem{razaghpanah2018apps}
Abbas Razaghpanah, Rishab Nithyanand, Narseo Vallina-Rodriguez, Srikanth
  Sundaresan, Mark Allman, Christian Kreibich, and Phillipa Gill.
\newblock Apps, trackers, privacy, and regulators: A global study of the mobile
  tracking ecosystem.
\newblock In {\em Network and Distributed System Security (NDSS)}, 2018.

\bibitem{sarabi2018characterizing}
Armin Sarabi and Mingyan Liu.
\newblock Characterizing the internet host population using deep learning: A
  universal and lightweight numerical embedding.
\newblock In {\em Proceedings of the Internet Measurement Conference 2018},
  pages 133--146, 2018.

\bibitem{wang2014whowas}
Liang Wang, Antonio Nappa, Juan Caballero, Thomas Ristenpart, and Aditya
  Akella.
\newblock Whowas: A platform for measuring web deployments on iaas clouds.
\newblock In {\em Proceedings of the 2014 Conference on Internet Measurement
  Conference}, pages 101--114, 2014.

\bibitem{le2018urlnet}
Hung Le, Quang Pham, Doyen Sahoo, and Steven C.~H. Hoi.
\newblock Urlnet: Learning a url representation with deep learning for
  malicious url detection, 2018.

\bibitem{bell2020analysis}
Simon Bell and Peter Komisarczuk.
\newblock An analysis of phishing blacklists: Google safe browsing, openphish,
  and phishtank.
\newblock In {\em Proceedings of the Australasian Computer Science Week
  Multiconference}, pages 1--11, 2020.

\bibitem{li2019stacking}
Yukun Li, Zhenguo Yang, Xu~Chen, Huaping Yuan, and Wenyin Liu.
\newblock A stacking model using {URL} and {HTML} features for phishing webpage
  detection.
\newblock {\em Future Generation Computer Systems}, 94:27--39, May 2019.

\bibitem{jeosandbox}
JoeSandboxCloud.
\newblock Analysis report ubar-pro4[.]ru.
\newblock \url{https://www.joesandbox.com/analysis/229408/0/html}.
\newblock Accessed: 2021-05-31.

\bibitem{sophos}
Sophos News.
\newblock Nearly a quarter of malware now communicates using tls.
\newblock
  \url{https://news.sophos.com/en-us/2020/02/18/nearly-a-quarter-of-malware-now-communicates-using-tls/}.
\newblock Accessed: 2021-05-31.

\end{thebibliography}

\appendix

\section{Ethical Considerations}
This work does not raise any ethical issues. The obtained dataset purports to study Internet measurement in good faith, and is secured privately with access granted only to the authors' affiliations. The dataset does not include any personally identifiable information, with all examples in this work anonymised so as to prevent linking sensitive activities to specific organizations or users.

\end{document}